\documentclass{emulateapj}

\usepackage{natbib}
\begin{document}
\title{MN Lup: X-rays from a weakly accreting T~Tauri star\thanks{Based on observations made with \emph{XMM-Newton}, an ESA science mission with instruments and contributions directly funded by ESA Member States and NASA, and ESO Telescopes at the La Silla Paranal Observatory under program ID 087.V0991(A).}}

\author{H.~M.~G\"unther}
\affil{Harvard-Smithsonian Center for Astrophysics, 60 Garden Street, Cambridge, MA 02138, USA}
\email{hguenther@cfa.harvard.edu}

\and

\author{U.~Wolter, J.~Robrade}
\affil{Universit\"at Hamburg, Hamburger Sternwarte, Gojenbergsweg 112, 21029 Hamburg, Germany}

\and

\author{S. J. Wolk}
\affil{Harvard-Smithsonian Center for Astrophysics, 60 Garden Street, Cambridge, MA 02138, USA}

\begin{abstract}
Young T Tauri stars (TTS) are surrounded by an accretion disk, which
over time disperses due to photoevaporation, accretion, and possibly
planet formation. The accretion shock on the
central star produces an UV/optical veiling continuum, line emission,
and X-ray signatures. As the accretion rate decreases, the impact on
the central star must change. In this article we study MN~Lup, a young
star where no indications of a disk are seen in IR observations. We present \emph{XMM-Newton} and
\emph{VLT}/UVES observations, some of them taken
simultaneously. The X-ray data show that MN~Lup is an active star
with $L_X/L_{bol}$ close to the saturation limit. However, we find high densities ($n_e > 3\times
10^{10}$~cm$^{-3}$) in the X-ray grating spectrum. 
This can be well fitted using an accretion shock model with an
accretion rate of $2\times10^{-11}M_{\sun}$~yr$^{-1}$.
Despite the simple H$\alpha$ line profile which has a broad component, but no absorption
signatures as typically seen on accreting TTS, we find rotational
modulation in \ion{Ca}{2}~K and in photospheric absorption lines. 
In
the H$\alpha$ line we see a prominence in absorption about $2R_*$
above the stellar surface - the first of its kind on a TTS.
MN~Lup is also the only TTS where accretion is seen, but no dust
disk is detected that could fuel it. We suggest that MN~Lup presents a
unique and short-lived state in the disk evolution. It may have lost
its dust disk only recently and is now accreting the remaining gas at
a very low rate.

\end{abstract}


\section{Introduction}
\label{sect:intro}
T~Tauri stars (TTS) are pre-main sequence stars of spectral type M to F
surrounded by a circumstellar disk. 
They come in two flavors, the classical T~Tauri stars (CTTS), which
actively accrete from their disk, and the more evolved weak-lined
T~Tauri stars (WTTS) whose accretion has already stopped. The disks are
typically detected in the IR, e.g.\ in \emph{Spitzer} surveys where
they show up as excess emission above the stellar photosphere. To
distinguish between CTTS and WTTS, the combination of strong H$\alpha$
emission and line asymmetry can be used because these are
reliable accretion indicators
\citep{1998ApJ...492..743M,2003ApJ...592..266M}. However, there are
some exceptions where a low H$\alpha$ equivalent width (EW) seems to correspond to high
mass accretion rates \citep{2004MNRAS.347..937L}; furthermore, the
strength of the intrinsic H$\alpha$ line also depends on the spectral
type and on the chromospheric activity of the specific star. In
extreme cases, the chromosphere alone can produce H$\alpha$ in
emission with an EW of a few \AA{}.

Several pathways are possible to evolve an accreting CTTS to a non-accreting WTTS through disk evolution. \citet{2012ApJ...747..103E} differentiate between ``pre-transitional disks'', where a gap in the dust disk is opened at some radius and ``transitional disks'' \citep{1989AJ.....97.1451S} where the innermost 10-50~AU are entirely cleared of micron sized dust grains. Gaps in disks are particularly interesting, since they indicate planet formation -- a planet located in the disk would clear its surroundings \citep{1980ApJ...241..425G,2003MNRAS.342...79R,2004ApJ...612L.137Q} and interrupt the mass accretion. Still, some dust often remains inside of the inner disk wall \citep{2010ApJ...717..441E}. Sufficiently close-in planets could also be responsible for an inner hole, but other effects, such as photo-evaporation \citep{2001MNRAS.328..485C,2009ApJ...690.1539G,2012MNRAS.422.1880O} or the magneto-rotational instability (MRI) \citep{2007NatPh...3..604C}, can cause an inside-out clearing as well.

All TTS have been known for a long time to be copious X-ray emitters \citep[see reviews by][]{1999ARA&A..37..363F,2012arXiv1210.4182G}. Particularly, they exhibit high levels of coronal activity and stellar flares.
Interferometric observations and modeling of the spectral energy distributions reveal that even in CTTS the accretion disk does not reach down to the star, but is truncated at a few stellar radii by the stellar magnetic field. X-ray and FUV radiation ionize the material at the inner disk rim. Thus, the accreting matter follows the magnetic field lines and impacts onto the stellar surface with a velocity close to free-fall \citep{1991ApJ...370L..39K,1994ApJ...429..781S}. For a dipolar field the impact would be at high stellar latitude, but modern simulations explore more complicated geometries \citep[e.g.][]{2012NewA...17..232L}.

The kinetic energy of the infalling matter is released in a standing shock wave, which heats the gas to temperatures of the order of 2-3~MK. This accretion hot spot explains, among other things, the optical veiling of TTS \citep{calvetgullbring}. 

With sufficiently high quality X-ray data CTTS can be singled out from other X-ray sources by, first, their strong soft X-ray excess \citep{RULup,manuelnh,2011AN....332..448G} and, second, by unusually high densities in the He-like triplets first seen in the CTTS \object{TW Hya} \citep{2002ApJ...567..434K}. These densities were confirmed later in several consecutively deeper observations \citep{twhya,2009A&A...505..755R,2010ApJ...710.1835B}. High densities in the X-ray emitting regions have been found in a number of CTTS since: \object{BP Tau} \citep{bptau}, \object{V4046~Sgr} \citep{v4046,2012arXiv1204.0964A}, \object{RU Lup} \citep{RULup}, \object{MP Mus} \citep{2007A&A...465L...5A}, \object{Hen 3-600} \citep{2007ApJ...671..592H}, and \object{V2129 Oph} \citep{2011A&A...530A...1A}. However, T~Tau itself, although known for its high accretion rate, is an exception to this rule \citep{ttau}.
The high densities observed in the He-like triplets can be naturally linked to accretion. The same model nicely explains the observed soft excess and allows us to reconstruct the density and velocity of the infalling material from X-ray emission lines \citep{acc_model}.
Studies in different wavelength regions regularly indicate very different sizes for the accretion spot \citep{acc_model,2011A&A...526A.104C}. The reason is unclear, but one explanation would be an inhomogeneous accretion spot, where only parts reach the temperatures of X-ray emitting gas; another possibility would be that that the accretion powers an enhanced chromosphere and corona, that emits locally around the accretion shock \citep{2010ApJ...710.1835B,2011A&A...535A...6P}. Also, the models often assume optically thin emission regions. This depends critically on the dimensions and shape of the accretion spot, but we do not know if the accretion is concentrated in a large, single spot or more widely distributed. 

We observed the TTS \object{MN Lup} simultaneously with \emph{XMM-Newton} and the \emph{VLT}. In this paper, we concentrate on the results of the X-ray observations. In Sect.~\ref{sect:stellarproperties} we summarize the stellar properties of MN~Lup, Sect.~\ref{sect:obs} contains the details of the observations and the data reduction. Sect.~\ref{sect:results} gives the results and Sect.~\ref{sect:discussion} a short discussion. We end with a summary (Sect.~\ref{sect:summary}).

\section{Stellar properties of MN Lup}
\label{sect:stellarproperties}
MN~Lup was discovered in an X-ray survey of the Lupus star forming region located at a distance of about 150~pc \citep{1997A&AS..123..329K}. Table~\ref{tab:mnlup} shows the stellar parameters. 
\begin{deluxetable}{lcr}
\tablecaption{Stellar parameters from \protect{\citet{2005A&A...440.1105S}}\label{tab:mnlup}}
\tablewidth{0pt}
\tablehead{
\colhead{parameter} & \colhead{symbol} & \colhead{value}}
\startdata
distance & $d$ & 150 pc\\
spectral type & & M0\\
luminosity & $L_*$ & $0.15 \pm0.03~L_\odot$\\
radius & $R_*$ & $0.90\pm0.02R_{\sun}$\\
mass\tablenotemark{a}  & $M_*$ & $0.68^{+0.06}_{-0.10} M_\odot $\\
mass\tablenotemark{b}  & $M_*$ & $0.54\pm0.12~M_\odot$\\
age   &    & 20 $\pm $ 10 Myr\\
period & $P$ & $0.439\pm0.005$ days\\
$v\sin i$ & & $74.6\pm1$~km~s$^{-1}$\\
inclination & $i$ & $40-50^\circ$
\enddata
\tablenotetext{a}{evolutionary tracks by \protect{\citet{1998A&A...337..403B}}}
\tablenotetext{b}{evolutionary tracks by \protect{\citet{2000A&A...358..593S}}}
\end{deluxetable}

MN~Lup is a pre-main sequence star as evidenced by its Li abundance. 
\emph{VLT}/UVES Doppler imaging over more than two rotations reveals structures which \citet{2005A&A...440.1105S} interpret as hot spots at high stellar latitude -- exactly where they are expected according to the magnetically funneled accretion theory. Very close to them patches of lower intensity are seen. \citet{2005A&A...440.1105S} interpret this as partially shadowed stellar photosphere.  The H$\alpha$ EW is $6.73\pm0.52$~\AA{}, higher than in quiescent main-sequence M0 stars \citep{2006AJ....132..866R}, but lower than in accreting CTTS, indicating some stellar activity or a low accretion rate. The intermediate inclination angle ensures that MN~Lup is not obscured by its disk.

\citet{1997A&A...320..185W} derive an optical reddening of $A_V=1.11$. However, in CTTS the veiling changes the intrinsic colors. \citet{1998ApJ...492..323G} therefore suggest to use the $V-R$ color, which is less affected. With this method \citet{2005A&A...440.1105S} derive $A_V=0.78$. Based on the higher extinction and a lower distance to MN~Lup of only 140~pc \citet{1997A&AS..123..329K} derive an X-ray luminosity of $L_X = 8.6\times10^{29}$~erg~s$^{-1}$ from \emph{ROSAT} observations.

MN~Lup has been observed with \emph{Spitzer} in the c2d survey \citep{2006ApJ...645.1283P} using the IRAC (3.6-8.0$\mu$m) and MIPS (24$\mu$m, no detection at 70$\mu$m) instruments. No IR excess above a stellar photosphere, which would indicate the presence of a dusty disk, was found down to a limit of a fractional disk luminosity $L_d/L_* < 9.1 \times 10^{-4}$, significantly lower than for the debris disk of $\beta$~Pic with $L_d/L_* = 2 \times 10^{-3}$ \citep{2010ApJ...724..835W}. The c2d survey is sensitive to the dust in disks from a few tenths of an AU out to a few tens of AU.

\section{Observations and data reduction}
\label{sect:obs}
\begin{deluxetable}{lccc}
\tablecaption{Log of \emph{XMM-Newton} observations\label{tab:obslog}}
\tablewidth{0pt}
\tablehead{
\colhead{ObsID} & \colhead{duration} & \colhead{date} & \colhead{OM}}
\startdata
0655760101 & 27 ks & 2010-08-03 & UVW1\\
0670580101 & 39 ks & 2011-08-12 & UVW1\\
0670580201 & 43 ks & 2011-08-12 & UV grism
\enddata
\end{deluxetable}
MN~Lup was observed three times with the X-ray satellite \emph{XMM-Newton}, which provides simultaneous data from the three imaging cameras PN, MOS1 and MOS2 and two reflection grating spectrometers (RGS1 and RGS2). In addition, there is an optical monitor (OM), which provides photometry or grism spectroscopy in the UV to optical range. In 2010 we performed a relatively short exposure to verify the \emph{ROSAT} count rate and test the spectral shape of the source. In 2011 we obtained observations with \emph{XMM-Newton} mostly simultaneous to optical spectroscopy from \emph{VLT}/UVES. This required splitting the X-ray data in two observations in 2011. In this article we concentrate on the X-ray data. 
Table~\ref{tab:obslog} summarizes the \emph{XMM-Newton} observations. In all cases we used the medium filter to reduce the contamination with optical light. All observations are processed with the Science Analysis System (SAS), version 11.0.0 \citep{2004ASPC..314..759G}. We merged all three EPIC detectors for the lightcurve analysis. In all exposures the extraction regions are small (30\arcsec{} in the MOS and 15\arcsec{} in the PN), thus the background is negligible compared to the source signal for lightcurves in the energy range 0.2-5.0~keV. For the spectral analysis we applied standard selection criteria to remove phases of high proton background, which would significantly increase the background for high energies. Specifically, the PN observation in 2010 is contaminated, so we use the MOS1 and MOS2 detectors only for spectral fitting in this exposure. For the exposures in 2011 PN, MOS1 and MOS2 data is fitted simultaneously.

Spectral models are fitted using the SHERPA tool \citep{2007ASPC..376..543D}. Line fluxes are modeled by intrinsically narrow Gaussian profiles, which are then convolved with the instrumental response function. In the \ion{Ne}{9} and \ion{O}{7} triplets all three lines are fitted simultaneously and the difference in wavelength is fixed. We allow for an overall adjustment of the wavelength to correct a possible error of the wavelength zero point. In all cases, the fitted wavelength agrees with the theoretical value within the uncertainties. In addition to the lines, we fit a constant to account for instrumental background and source continuum. Because the width of the fit window for each line extends only 0.2~\AA{} on either side of the line center this is a good approximation.

The \emph{VLT}/UVES data (Program ID 087.C-0991(A)) was taken on the nights starting 2011-08-11 (13 exposures) and 2011-08-12 (12 exposures), details on the instrument can be found in \citet{2000SPIE.4008..534D}. The exposure time was set to 20~minutes throughout. UVES was operated in dichroic mode. The slit width in both UVES arms was 0.9\arcsec, resulting in a spectral resolution of about 45000. The UVES data was reduced using the ESO pipeline .
The relative timing of the \emph{XMM-Newton} and \emph{VLT}/UVES observations can be seen in Fig.~\ref{fig:lc}.

\section{Results}
\label{sect:results}

\subsection{Lightcurves}
\begin{figure*}
\plotone{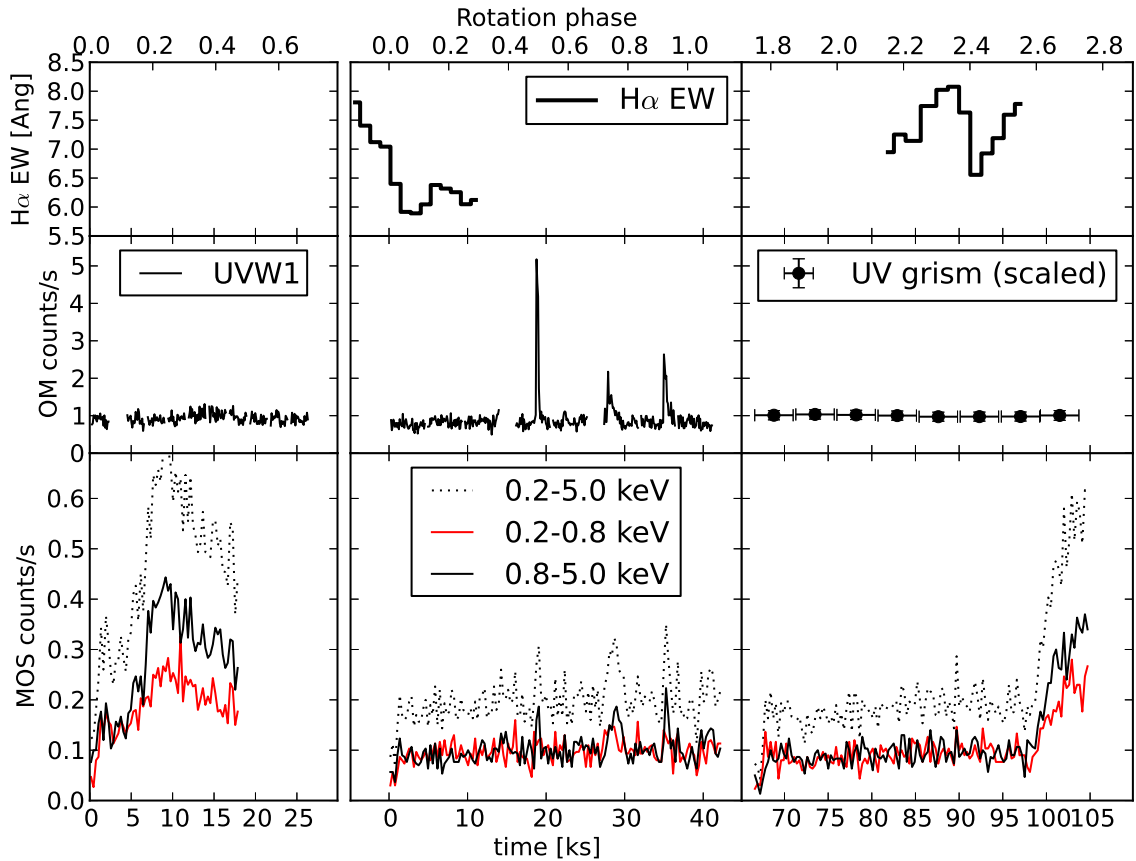}
\caption{Lightcurves from all three \emph{XMM-Newton} observations and H$\alpha$ EW from \emph{VLT}/UVES. The first X-ray observation is from 2010-08-03, the other two are taken about one year later on 2011-08-12. The time is counted from the beginning of the first X-ray observation in each year. Similarly, the rotational phase is set to 0 at $t=0$. \emph{Top}: H$\alpha$ EW from the UVES data. \emph{Middle}: OM data, binned to 100~s. In the last observation the OM operated with the UV grism, which has a different wavelength range. The rate is scaled to match the quiescent UVW1 count rate.  Error bars along the x-axis indicate the exposure time of the grism data. \emph{Bottom}: EPIC count rate, binned to 300~s. In the energy bands shown the background is negligible.  See the electronic edition of the Journal for a color version 
of this figure.\label{fig:lc}}
\end{figure*}

In this section we discuss the lightcurves extracted from the CCD detectors and the OM. In figure \ref{fig:lc}, time is counted from the beginning of the first  observation in each year. The rotation period of MN~Lup is $0.0439\pm0.005$~days \citep{2005A&A...440.1105S}. This uncertainty amounts to 10 rotations over one year, which is too large to calculate the relative phase between the observations taken in 2010 and 2011. Thus, the rotational phase is arbitrarily set to 0 at the beginning of the first X-ray observation in each year.

Table~\ref{tab:obslog} shows that the OM operated with the UVW1
filter, which approximately covers the wavelength range
2500-3500~\AA{} in the first two observations and with the UV grism in
the last observation. The middle panels in Fig.~\ref{fig:lc} show the
OM lightcurve. In the first two observations, the count rate of MN~Lup
was read out in a fast window, which provides very high time
resolution. However, MN~Lup is so bright, that within the fast window
little background area is available. Thus, the background subtraction
is very sensitive to an accurate positioning of the target. This leads to small changes in the quiescent count rate between individual exposures. During one X-ray observation, the OM performed multiple exposures. Time delay between exposures causes missing points in the lightcurve, e.g. at 15 and 28~ks in the second observation. The luminosity of MN~Lup is comparable to other sources in the field, thus its dispersed grism spectrum is contaminated by the zeroth order of several field stars. Therefore, we extract only the zeroth order of our target MN~Lup using simple aperture photometry on the pipeline processed OM images. While the count rate in the zeroth order is not flux calibrated, we should be able to see relative changes in brightness during the observation. The UV grism covers the wavelength range 1800-3500~\AA{} and its count rates are scaled to match the observations with the UVW1 filter at the quiescent level.

The lower panels of Fig.~\ref{fig:lc} show the X-ray lightcurve. It is divided into two bands from 0.2-0.8~keV and 0.8-5.0~keV. There is little signal above 5~keV and the separation between the two bands is chosen such that their count rates (and thus the signal-to-noise ratio (SNR) in the lightcurves) are comparable.
The upper panels in the figure show the H$\alpha$ EW, which is discussed in Sect.~\ref{sect:halpha}. 

The UV lightcurve is mostly flat around 1~count~s$^{-1}$ with three
noticeable flares, all in the second observation around 19, 28 and
36~ks; the corresponding increases in count rate are factors of 5, 2,
and 2.5, respectively. Those flares are very short (about 0.5, 1.5 and
1~ks) and all of them are seen in X-rays, most clearly in the hard band. The largest change in X-ray count rate is observed for the last flare, where the count rate doubles. The brightest UV flare, at 19~ks, has only a weak X-ray counterpart. Conversely, the X-ray count rate was much higher in 2010 (first observation) than in the quiescent state in 2011, but the OM count rate is almost the same; the small difference might be due to the difficulties in the background determination in the fast window. Similarly, the X-ray count rate increases in the last observation after about 100~ks by a factor of 2 and shows notable hardening, but no comparable change is seen in the OM.

The hardening of the X-ray emission during the three short flares
in the second observation suggests that these are caused by coronal
activity as opposed to accretion events. However, it is then
surprising that the much longer X-ray activity in the first and third
observation has no or only weak UV counterparts. Correlations between
X-ray and UV/optical emission in CTTS have been searched for for some
time \citep[][]{1997A&A...324..155G} with little success. In
surveys of the ONC \citep{2006ApJ...649..914S} and the Taurus molecular cloud \citep{2007A&A...468..379A} X-ray and UV/optical lightcurves
are uncorrelated with very few exceptions. Both studies conclude, that
the X-ray variability is dominated by coronal activity, while the UV
shows the rotational modulation of accretion
spots. \citet{2006ApJ...649..914S} find two examples of short (hours)
X-ray and optical bursts, that they interpret as ``white-light''
flares. Flares are observed with a higher time resolution on non-accreting stars, such
as \object{CN Leonis} \citep{2008A&A...481..799S} and \object{Proxima
  Centauri} \citep{2011A&A...534A.133F}. Both studies used
simultaneous \emph{XMM-Newton} and \emph{VLT}/UVES data in a similar setup to
our observations of MN~Lup and find bright
flares (increase in luminosity $> 10$ in X-ray, UV, and optical) that last less than an
hour. The higher luminosity in those observations allows the authors
the constrain the physics better and they show that the events are due
to coronal and chromospheric activity. The three short bursts in our
second observation show a similar pattern and we conclude that we observe three short
``white-light'' flares on MN~Lup.

\subsection{CCD spectroscopy}
\label{sect:lowres}
We fit the CCD spectra of all three observations simultaneously. The data are binned to 15 counts per bin. The fit is done with a single absorption component and three optically thin, collisionally dominated emission components \citep[APEC models,][]{2005AIPC..774..405B}. We couple the temperatures of the three components, such that the value of $T_1$, $T_2$, and $T_3$ is the same in each exposure, but allow independent normalizations for 2010, 2011 quiescent phase ($t < 98$~ks), and 2011 X-ray flare ($t> 98$~ks) because the lightcurve already shows that MN~Lup was significantly brighter in 2010 than in 2011. This model gives a statistically acceptable fit (red. $\chi^2 = 0.93$), but all spectra show systematic residuals at energies where strong emission lines are found in the grating spectrum. Thus, we also fit the abundances of the most important elements (O, Ne, Fe); table~\ref{tab:EPIC} gives the best-fit values relative to the solar abundances of \citet{1998SSRv...85..161G}. The addition of the abundances as free parameters, which is \emph{not required} by the $\chi^2$ value, but motivated from systematics in the residuals and the grating spectrum, leads to a $\chi^2 \ll 1$ and larger uncertainties on all parameters of the model. The abundances show a trend with the inverse first ionization potential (IFIP effect), where elements like Ne with a high first ionization potential are more abundant compared with the sun and elements of low FIP like Fe are less abundant. Such an abundance pattern is typical for active stars \citep[see review of ][]{2009A&ARv..17..309G}.

\begin{figure}
\plotone{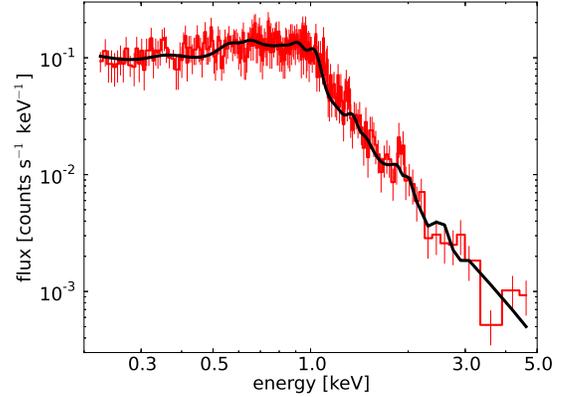}
\caption{PN spectrum from the second observation binned to 15 counts per bin (red/gray). The best fit model with three optically thin, collisionally dominated emission components is overlaid (black). See the electronic edition of the Journal for a color version 
of this figure. \label{fig:PN}}
\end{figure}

In CCD spectroscopy there is often an ambiguity between absorption and the amount of cool plasma, such that a higher absorbing column density can be compensated by a larger emission measure with very little changes in the total spectrum; this ambiguity can only be broken with grating spectroscopy of high SNR. Therefore we consider two cases: First, we fix the absorption to $N_H = 1.4\times10^{21}$~cm$^{-2}$, which corresponds to $A_V = 0.78$ \citep{2005A&A...440.1105S} using the relation between optical reddening and X-ray column density from \citet{2003A&A...408..581V}. In the second model we fit the absorption as a free parameter, resulting in $N_H = (2\pm1)\times10^{20}$~cm$^{-2}$ (table~\ref{tab:EPIC}). Figure~\ref{fig:PN} displays the model and the PN data from the second observation.
The fit values for both models agree within the errors except for the volume emission measure ($VEM$) of the soft component, which is about 25 times larger for the fixed absorbing column density of $N_H = 1.4\times10^{21}$~cm$^{-2}$. 

In previous Doppler imaging, large-scale temperature differences are found over the stellar photosphere. These variations are interpreted as absorption by the accretion column \citep{2005A&A...440.1105S}. If this interpretation is correct, then the X-ray emission might or might not be subject to the same absorbing column density, depending on the location of the emitting region. If the coronal structures are located close to the pole, then they could be seen through significantly less absorption. X-ray emission from the accretion shock might be partly covered by the accretion funnels, but in this case we would expect significant absorption in the fit. 
Consequently, the unabsorbed luminosities $L_X$ differ by factors of a few between the two models. \citet{2005A&A...440.1105S} calculate the bolometric luminosity of MN~Lup as $L_{bol} =  0.14\pm 0.02 L_{\sun}$, which leads to $\log(L_X/L_{bol}) = -2.3$ or $-2.7$, for the high or low value of $N_H$, respectively. Cool stars saturate at an activity level of $\log(L_X/L_{bol}) \approx -3$ \citep[][and references therein]{1994ApJS...91..625S,2009A&ARv..17..309G}. In either case MN~Lup is an active star consistent with the IFIP pattern found in the abundances. However, if the model with the higher value for $N_H$ were correct, then MN~Lup would be significantly brighter than the saturation limit and its X-ray flux would be totally dominated by the cool plasma. This would imply high values for the accretion rate, which are incompatible with the H$\alpha$ data (see Sect.~\ref{sect:halpha}). Thus, we consider the model with $N_H = 1.4\times10^{21}$~cm$^{-2}$ unphysical. 

Comparing the flare and the quiescent phase in 2011, we see that the volume emission measure in the flare increases significantly. The strongest increase, by a factor of five, happens in the hottest component. This is not unexpected. Given that the temperature $T_3$ in our model is forced to be the same for all exposures, the extra flare emission must manifest itself as a larger VEM. The $VEM$ in 2010 is comparable to the flare phase in 2011 for all temperature components, which indicates that the observation in 2010 happened in a period of considerable activity as well.

\begin{table}
\caption{\label{tab:EPIC} Best-fit model parameters (1$\sigma$ confidence intervals).}
\begin{center}
\begin{tabular}{llrrr}
\hline \hline
parameter & unit & 2010 & 2011 & 2011 \\
 &  & MOS only & quiescent & flare \\
\hline
$N_H$ & $10^{21}$ cm$^{-2}$ & \ldots & $0.2\pm0.1$\tablenotemark{c} & \ldots\\
$VEM_1$ & $10^{52}$ cm$^{-3}$ &$2\pm1$   &$1.3^{+0.7}_{-0.2}$ & $1.9_{-0.7}^{+1.5}$\\
$VEM_2$ & $10^{52}$ cm$^{-3}$ &$3.8\pm0.6$ & $2.4\pm0.3$   & $3.1\pm0.6$\\
$VEM_3$ & $10^{52}$ cm$^{-3}$ &$12.5\pm0.6$ & $2.6\pm0.2$   &$13.3\pm0.9$\\
$kT_1$ & keV & \ldots & $0.14^{+0.02}_{-0.01}$\tablenotemark{c} & \ldots\\
$kT_2$ & keV & \ldots & $0.43^{+0.04}_{-0.01}$\tablenotemark{c} & \ldots\\
$kT_3$ & keV & \ldots & $2.3\pm0.2$\tablenotemark{c} & \ldots\\
$\log L_X$\tablenotemark{a}  & erg s$^{-1}$ & 30.4 & 30.0 & 30.4 \\
O  & \tablenotemark{b} & \ldots & $0.69^{+0.15}_{-0.08}$\tablenotemark{c} & \ldots\\
Ne & \tablenotemark{b} & \ldots & $2.2\pm0.3$\tablenotemark{c} & \ldots\\
Fe & \tablenotemark{b} & \ldots & $0.45\pm0.07$\tablenotemark{c} & \ldots\\
\hline
red. $\chi^2$ & \multicolumn{4}{c}{0.71 (for simultaneous fit of all 3 datasets)}\\
\hline
\end{tabular}
\end{center}
\tablenotetext{a}{unabsorbed flux in the energy range 0.2-5.0 keV}
\tablenotetext{b}{relative to \citet{1998SSRv...85..161G}}
\tablenotetext{c}{This value is used for all three datasets.}
\end{table}

\subsection{X-ray grating spectroscopy}
\label{sect:gratings}
\begin{figure*}
\plotone{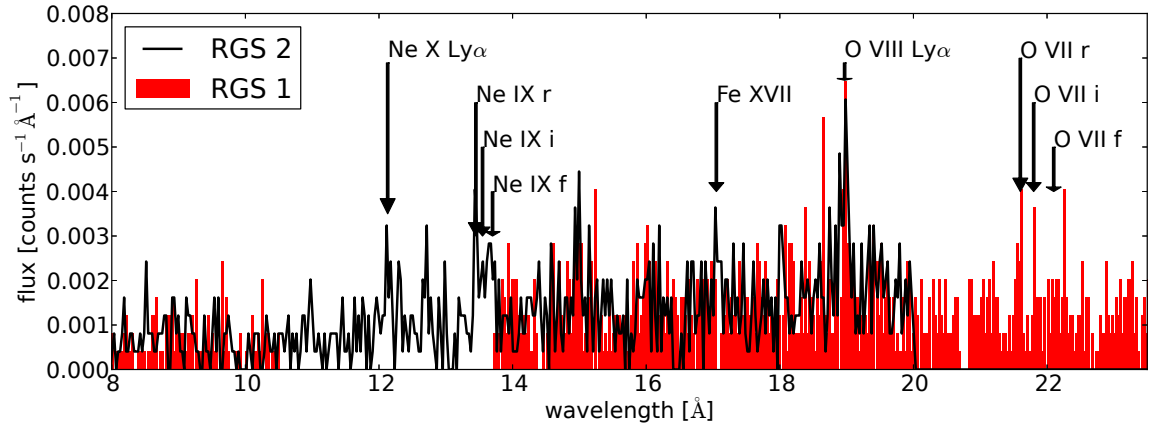}
\caption{RGS spectra of MN Lup. See the electronic edition of the Journal for a color version of this figure. \label{fig:RGS}}
\end{figure*}

Figure~\ref{fig:RGS} shows the merged RGS spectra for the observations in 2011. The signal in the high-resolution grating spectrum is low, thus only few line fluxes can be measured. 
All lines strong enough to fit the line flux are marked and given in table~\ref{tab:RGS}. The intrinsic line width is much less than the instrumental broadening and thus can be fixed. From the fitted number of counts and the effective area the intrinsic line luminosity is calculated assuming a distance of 150~pc and the absorbing column density from table~\ref{tab:EPIC}; the uncertainty of the count number and the uncertainty on $N_H$ both contribute to the uncertainties on the luminosity given in the table. The detected lines originate from H and He-like neon and oxygen and from \ion{Fe}{17}. 
Due to missing chips the wavelength coverage of both RGS is not
complete and \ion{O}{8} 18.97~\AA{} is the only line detected in both
detectors. Its flux is compatible in RGS1 and RGS2.

\begin{table}
\caption{\label{tab:RGS} Line fluxes  (1$\sigma$ confidence intervals).}
\begin{center}
\begin{tabular}{lrrrr}
\hline \hline
Line & $\lambda$ & counts & photon flux & luminosity\tablenotemark{a} \\
     & [\AA{}]   &        & [s$^{-1}$ cm$^{-2}$]&   [$10^{27}$ erg s$^{-1}$]\\
\hline
\multicolumn{4}{c}{RGS1}\\
\hline
\ion{O}{8} Ly$\alpha$  & 18.97 & 38.8 & $11.4\times10^{-6}$ &$37\pm9$\\
\ion{O}{7} r  & 21.6  & 16.7 & $5.8\times10^{-6}$& $17\pm7$\\
\ion{O}{7} i  & 21.8  & 11.7 & $4.3\times10^{-6}$& $13\pm6$\\
\ion{O}{7} f  & 22.1  &  8.9 & $3.2\times10^{-6}$& $10\pm6$\\

\hline
\multicolumn{4}{c}{RGS2}\\
\hline
\ion{Ne}{10} Ly$\alpha$ & 12.13 & 14.3 & $3.1\times10^{-6}$& $14\pm9$\\
\ion{Ne}{9} r & 13.45 & 21.6 & $4.6\times10^{-6}$& $19\pm8$\\
\ion{Ne}{9} i & 13.55 & 12.0 & $2.5\times10^{-6}$& $11\pm7$\\
\ion{Ne}{9} f & 13.70 & 21.2 & $4.4\times10^{-6}$& $19\pm9$\\
\ion{Fe}{17}  & 17.05/17.10 & 20.9 & $5.0\times10^{-6}$& $18\pm11$\\ 
\ion{O}{8} Ly$\alpha$  & 18.97 & 39.1 & $10.2\times10^{-6}$& $33\pm10$\\

\hline
\end{tabular}
\end{center}
\tablenotetext{a}{corrected for $N_H=2.1\times10^{20}$~cm$^{-2}$}
\end{table}

The He-like triplets of \ion{Ne}{9} and \ion{O}{7} are particularly important, because their line ratios are temperature and density-sensitive. These triplets consist of a resonance ($r$), an intercombination ($i$), and a forbidden line ($f$) \citep{1969MNRAS.145..241G,2001A&A...376.1113P}. Two ratios ($R = f/i$ and $G = (f+i)/r$) are defined from the three observed lines. If the emitting plasma has high electron densities $n_e$ or strong ambient UV photon fields then the $R$-ratio falls below its low-density limit, because electrons are collisionally or radiatively excited from the upper level of the $f$ to the $i$ line. In MN~Lup the UV field is insufficient to influence the $R$-ratio of the observed Ne and O triplets, so any deviation must be caused by the density in the emitting region. The $G$-ratio varies slowly with the plasma temperature. The low count numbers and consequently large uncertainties do not allow us to constrain the temperature meaningfully from this ratio.

\begin{figure}
\plotone{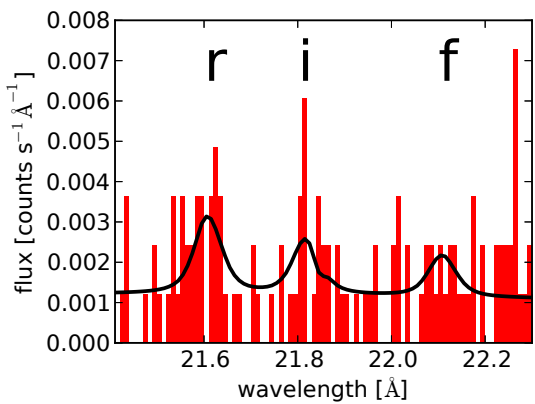}
\caption{\ion{O}{7} He-like triplet with best-fit model. The position of the recombination ($r$), intercombination ($i$) and forbidden ($f$) lines are marked. The best-fit clearly deviates from the low-density regime where $f/i=3.8$. \label{fig:o7}}
\end{figure}

Figure~\ref{fig:o7} shows the \ion{O}{7} triplet. The measured
$R$-ratio is 0.75 ($n_e \approx 2\times 10^{11}$~cm$^{-3}$). Due to
the low number of counts the errors are non-Gaussian and strongly
asymmetric, so we performed Monte-Carlo simulations to estimate the
confidence limits. We run 10,000 simulations; in each of them the
number of counts in each bin is drawn from a Poisson distribution with
the best fit-model as expectation value. We fit the background level
and the amplitude of the three lines to the simulated spectra in the
same way we treat the observed data. This procedure is very similar to
that employed by \citet[][appendix A]{HD163296}, where more details
can be found. According to the CHIANTI spectral database \citep{CHIANTI,2009A&A...498..915D} the low-density limit of the \ion{O}{7} $R$-ratio is 3.8. Our Monte-Carlo simulations show that the data exclude the low-density limit ($n_e\lesssim 10^{9}$~cm$^{-3}$) with 93\% significance and give $1\sigma$ confidence for $R < 1.16$ ($n_e > 3\times10^{10}$~cm$^{-3}$), but we cannot place an upper limit on the density.

For \ion{Ne}{9} the measured $R$-ratio is 1.75 ($n_e \approx 4\times
10^{11}$~cm$^{-3}$). This triplet, especially the $i$ line, can be
blended by lines of \ion{Fe}{17}, \ion{Fe}{19} and \ion{Fe}{20} . Given the low signal-noise ratio in the data a detailed deblending as in \citet{2008MNRAS.385.1691N} is not possible. Using CHIANTI we find that the strongest blend from \ion{Fe}{17} is a line at 13.82~\AA{}; however, at its peak formation temperature, the flux in this line is still only 4\% of the flux in the \ion{Fe}{17} lines at 17.05 and 17.10~\AA{}. Thus, its maximum contribution to the \ion{Ne}{9} $f$ line is less then 10\% of the statistical uncertainty (table~\ref{tab:RGS}). To asses the importance of other Fe ionization stages, we simulate spectra using the emission measures in table~\ref{tab:EPIC}. Even in the flare, when the \ion{Fe}{19} and \ion{Fe}{20} lines are strongest, \ion{Fe}{20} 13.78 and 13.84~\AA{}, the strongest blends, have a flux of only 12\% of the flux in the \ion{Fe}{17} lines at 17.05 and 17.10~\AA{}. Since the emission measure of the hot plasma outside of the flare is only a fifth of the emission measure during the flare, the average contribution is much less. Thus, these lines also contribute less then 10\% of the statistical uncertainty to the \ion{Ne}{9} $f$ line. For the temperature observed, the contamination of the \ion{Ne}{9} $r$ and $i$ lines is even lower. Thus, we can neglect the blends and find that the observed ratio is compatible with the low-density limit ($n_e \lesssim 3\times 10^{10}$~cm$^{-3}$) on the $1\sigma$ level, but we can exclude densities above $3\times 10^{12}$~cm$^{-3}$ at the 95\% confidence level.

In summary, intermediate densities in the range $3\times 10^{10}<n_e < 3\times 10^{12}$~cm$^{-3}$ are compatible with both triplets. 
In the above discussion it is assumed that all emission seen in a
triplet originates in a homogeneous plasma. This is not
necessarily the case, as multiple components with different
densities could contribute to the observed line fluxes. To estimate
the minimum emission measure of the dense component, we consider the
most extreme case where the \ion{O}{7} triplet comes from a superposition of a
cool corona, which is in the low-density range, and some plasma in the
high-density limit. The corona would contribute virtually nothing to
the $i$ line, but explain most of the $f$ line flux, while the
converse is true for the dense component. Since both lines
are of comparable strength, each component must have about half of the
total emission measure for \ion{O}{7}.

\subsection{Soft excess}
\begin{figure}
\plotone{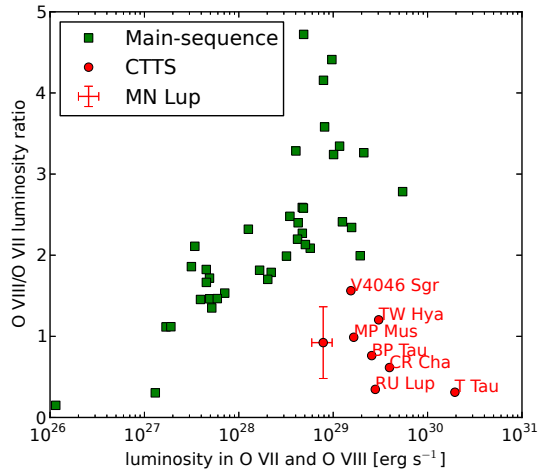}
\caption{The ratio of the \ion{O}{8} 18.97~\AA{} line and \ion{O}{7} triplet ($r+i+f$) luminosities is a temperature indicator. All CTTS show a soft excess when compared to MS stars of similar total luminosity. See the electronic edition of the Journal for a color version of this figure. \label{fig:o72o8}}
\end{figure}

CTTS show an excess of soft plasma compared to main-sequence (MS)
stars of similar total luminosity. Observationally, a lower ratio of
the  \ion{O}{8}/\ion{O}{7} luminosities \citep{RULup,manuelnh}
indicates an additional soft emission component around 1-2~MK. As this
is only seen in accreting CTTS and Herbig Ae/Be stars (HAeBe), it has
to be related to the accretion shock. Figure~\ref{fig:o72o8} shows the
\ion{O}{8}/\ion{O}{7} ratio for MS stars from the sample of
\citet{2004A&A...427..667N} and CTTS collected from the literature
(references are given in Sect.~\ref{sect:intro}). All fluxes are corrected for absorption using the $N_H$ values found in a global fit to the X-ray data.

In order to calculate the flux ratio, grating spectroscopy is required. Thus, only bright CTTS with little absorption can be placed on the diagram. The sample of MS stars includes much closer and fainter objects. When compared to MS stars of similar luminosity, CTTS show a soft excess. MN~Lup is shown in the figure as a CTTS, based on the fitted model with the lower absorption. It has the lowest total luminosity of all CTTS in the sample. The soft excess is almost certainly related to accretion, although it may not be formed in the post-shock zone.
\citet{2010ApJ...710.1835B} measure density differences between \ion{O}{7} and \ion{Ne}{9} in the CTTS TW~Hya and find that the \ion{O}{7} emission is not compatible with an origin in the post-shock cooling zone according to current shock models (see next section), but can be explained by a lower density corona fed by the accretion stream.

\subsection{Shock models}
\label{sect:shock}
The soft excess and the high densities found in the \ion{O}{7} triplet indicate an accretion shock origin of the soft plasma component. Therefore, we fit the CCD data and the RGS data simultaneously to an accretion shock model. Due to the low count number, the RGS data cannot be binned or the information contained in the He-like triplets would be lost. We use the Cash statistic \citep{1979ApJ...228..939C} as implemented in Sherpa, which is appropriate for Poisson-distributed low count data. Unlike the $\chi^2$ statistic, the Cash-statistic does not provide an intrinsic measure of the goodness-of-fit. We replace the coolest component of the three temperature fit in Sect.~\ref{sect:lowres} with the accretion shock models of \citet{acc_model}, updated in \citet{2011AN....332..448G}. Abundances and the absorbing column density are fixed at the values in table~\ref{tab:EPIC}. We fit the RGS data and the binned EPIC data from 2011. While we use all the RGS data, we restrict the EPIC data to the quiescent phase (before $t = 98$~ks, see Fig.~\ref{fig:lc}). The background for the EPIC detectors is negligible and we assume a constant background for the RGS, which is not subtracted from the data, but fitted simultaneously to conserve the Poisson counting statistic in the spectrum.

The shock model is described in detail in \citet{acc_model}.
The simulations assume a 1-dimensional geometry and an
optically thin post-shock cooling zone. The accreting matter is heated
in a strong shock and then cools down radiatively. Half of the
radiation is directed inwards and absorbed by deeper photospheric
layers, the other half of the radiation escapes. The temperature
distribution behind the shock is calculated and integrated spectra for
different pre-shock densities and velocities are tabulated. 
The model assumes a Maxwellian distribution for both electron and
ions, but it allows for different temperatures in those components,
since ions are heated to much higher temperatures in the shock and
Coulomb interactions between electrons and ions equilibrate the
temperatures only slowly. At each step, the simulation explicitly
solves the rate equations for ionization and recombination because
these time scales can be longer than the dynamical time scale for some
ions, so that the post-shock plasma is not in ionization
equilibrium. Thermal conduction is negligible in the post-shock
plasma, since the required time scales are much longer than the time it
takes for one packet of plasma to travel through the cooling
zone. Additionally, heat transport between the post-shock plasma and
the surrounding atmosphere is suppressed by the magnetic field of the
accretion funnel. 

The biggest caveat is that the simulations assume the medium to be
optically thin. No line opacity is expected in the X-ray range as
calculated a posterior and this is confirmed by the observed
$r/(f+i)$ ratio in the triplets which matches optically thin
predictions; we fit and account for continuum absorption $N_H$. However, \citet{2005CSSS.519.D}
argues that (a significant part) of the accretion shock could be
buried so deep in the atmosphere, that it is entirely absorbed. In
this case, the mass accretion rate that we fit below is to be
understood as a lower limit to the total mass accretion rate.

During the fit, the grid of spectra is interpolated in the parameters pre-shock velocity $v_0$ and density $n_0$. Table~\ref{tab:shock} shows the best-fit parameters of the shock model.

\begin{table}
\caption{\label{tab:shock} Best-fit model parameters for the quiescent interval in 2011 (1$\sigma$ confidence intervals).}
\begin{center}
\begin{tabular}{lrl}
\hline \hline
%
parameter & model & unit\\
\hline
$N_H$       & =0.21             & $10^{21}$cm$^{-2}$  \\
$VEM_1$     & $2.0\pm0.1$       & $10^{52}$ cm$^{-3}$ \\
$VEM_2$     & $2.1\pm0.2$       & $10^{52}$ cm$^{-3}$ \\
$kT_1$      & $0.77\pm0.04$     & keV                 \\
$kT_2$      & $2.4\pm0.2$       & keV                 \\
$v_0$       & $510 \pm50$       & km~s$^{-1}$         \\  
$n_0$       & $1\pm1$\tablenotemark{a}           & $10^{10}$ cm$^{-3}$ \\
$A_{spot}$  & $10_{-3}^{+1}$    & $10^{20}$ cm$^2$    \\
$\dot M$    & $2\times10^{-11}$ & $M_{\sun}$ yr$^{-1}$ \\
$L_{shock}/L_{corona}$  & 0.46  &        \\
\hline
\end{tabular}
\end{center}
\tablenotetext{a}{More stringent limits can be derived from analysis the He-like triplets alone (see text).}
\end{table}

Compared to the fit in Sect.~\ref{sect:lowres} the fit statistic
improves when one shock model and two single-temperature components
(which represent the coronal contribution) are used in the model
instead of three single-temperature components. This model reproduces
the \ion{O}{7} $R$-ratio. The emission measure and the
temperature of the hot components change only marginally. The shock
component has a pre-shock velocity $v_0$ of 500~km~s$^{-1}$ and a
pre-shock density $n_0$ of $1.0\times10^{10}$~cm$^{-3}$. The
statistical error on the velocity is smaller than the grid step size
of 50~km~s$^{-1}$. Given the accuracy of the model grid, we give the model grid step size in table~\ref{tab:shock} as uncertainty of the fit values. Due to
the low count number there is no formal lower limit on the density in the global fit. The normalization of the shock model is directly related to the shock area $A$. From these numbers the accretion rate $\dot M$ can be calculated as:
\begin{equation}
\dot M = \mu m_{\mathrm{H}} n_0 v_0 A\ ,
\end{equation}
where $\mu$ is the dimensionless mean particle mass, averaged over electrons and ions and $m_{\mathrm{H}}$ is the mass of the hydrogen atom.
Table~\ref{tab:shock} gives the mass accretion rate. While the
uncertainty on $n_0$ and $A$ are large, the
uncertainty on the mass accretion rate is much lower, because it is
determined from the normalization of the shock model, while the
density is only constrained by the $f/i$ ratio in the He-like
triplets. The fitted value for the size of the accretion spot implies
a surface filling factor of 2\%. As estimated in
Sect.~\ref{sect:gratings} at most half of the cool plasma can be
explained by coronal emission. Thus, the shock luminosity (and consequently
the accretion rate) of the shock may be overestimated by a factor of 2 at most.
The free-fall velocity from infinity is
\begin{equation}
v_{ff} = \sqrt{\frac{2GM_*}{R_*}}
\end{equation}
where $G$ is the gravitational constant. For a stellar radius of $R_* = 0.87R_{\sun}$ and a mass of $M_* = 0.68M_{\sun}$ \citep{2005A&A...440.1105S} this leads to a free-fall velocity of 550~km~s$^{-1}$ in very good agreement with the values derived above. 

\subsection{Variability in the optical spectrum}
The optical spectra of MN~Lup are dominated by the Balmer series and \ion{Ca}{2} H and K in
emission,  telluric absorption, and photospheric absorption
lines. MN~Lup has been Doppler imaged with \emph{VLT}/UVES before by
\citet{2005A&A...440.1105S} and we refer to that paper for a figure of
the optical spectrum of MN~Lup and its general properties. Our data is
very similar except for the absence of airglow emission that plagued
the analysis in the previous paper.  
\begin{figure}
\plotone{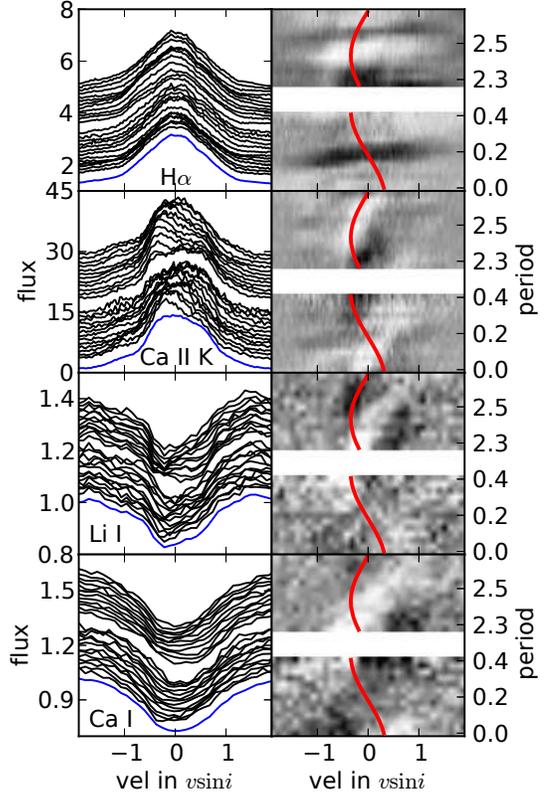}
\caption{Temporal variability of selected lines in MN~Lup
  (\ion{Ca}{2}~K at 3933.66~\AA{}, H$\alpha$ at 6564.89~\AA{},
  \ion{Li}{1} at 6707.84~\AA{}, and \ion{Ca}{1} at 6102.73~\AA{}).
  \emph{left}: The spectra are continuum normalized and offset along
  the y-axis for clarity. Time starts from the bottom and the break
  between the two nights is not to scale. Due to the lower signal, the
  absorption lines are binned to a larger bin width than the emission
  lines. The blue line in each panel shows the mean line
  profile. \emph{right}: Gray scale representation of the residuals to
  the mean spectrum. As an example, the red curve shows the position of a tentative spot with $\theta=70^{\circ}$ and phase $\phi = 0.68$. Clearly, the line deformations do not follow this red line through all observed phases, see text for discussion. \label{fig:lines}}
\end{figure}
Figure~\ref{fig:lines} shows the \ion{Ca}{2}~K, H$\alpha$, and
\ion{Li}{1} 6708~\AA{} lines and another absorption line that we
identify as \ion{Ca}{1} 6103~\AA{}. \citet{2005A&A...440.1105S}
suggest this to be the \ion{Fe}{1} line at 6102~\AA{} with a blend
from an excited Li line at 6104~\AA{}, but the VALD database
\citep{2000BaltA...9..590K} indicates that the \ion{Ca}{1} 
line is stronger given the effective temperature of MN~Lup and assuming solar metallicity.
Furthermore, this line identification agrees better with our
determined radial velocity of MN~Lup than Strassmeier et al.'s
identification.

Variability within a night and between the two nights can be seen in
all lines in the figure. All pronounced features move from blue to red
through the line profiles which is consistent with rotational
modulation. The features in \ion{Ca}{2}~K, \ion{Li}{1}, and
\ion{Ca}{1} are, at least partially, correlated and their slope in the
diagram is very similar. They are most clearly defined between phase
2.3 and 2.7. The second white structure in \ion{Ca}{2}~K between phase 0.0 and phase 0.2 is
less clear. 

The modulation features in the H$\alpha$ line profile differ strongly
from the other lines in the figure. Possibly, the bright \ion{Ca}{2}~K
feature around phase 2.5 has a broadened counterpart in H$\alpha$.
However, the most prominent variations in H$\alpha$ are the dark
features quickly passing through its profile around phases 0.2 and
2.6.  We tentatively interpret these dark bands as signatures of
co-rotating prominences. Such features are commonly seen on a few
highly active fast-rotating stars (most notably \object{AB Dor} and \object{BO Mic})
\citep{1996IAUS..176..449C,1999MNRAS.302..437D,2008A&A...478L..11W},
but, to our knowledge they have only been found on one other WTTS or
CTTS, namely \object{TWA 6} \citep{2008MNRAS.385..708S}.  However, the
spectral feature ascribed to a prominence in TWA~6 is only seen in
emission beyond the stellar disk and not in absorption above it
\citep[see Fig.~16 of][]{2008MNRAS.385..708S}. In our data of MN
Lup it is the other way around: We only see the tentative prominences
in absorption above the stellar disk.  The corresponding features pass from minus to plus
$v\sin i$ in about 0.1 rotations, which corresponds to a height of
about two stellar radii above the stellar surface.

Apart from the tentative prominence features, the H$\alpha$ profile
variations do not add any information to the following discussion. Thus, we concentrate on \ion{Ca}{2}~K and the two photospheric lines.

We do not attempt to model the line profiles of MN
Lup's ``undisturbed'' photosphere or chromosphere -- whatever
undisturbed could mean in detail in this context.  Instead, the gray
scale plots in Fig.~\ref{fig:lines} show the line variations relative
to the corresponding mean line profile obtained by averaging all our
observed spectra (qualitatively, the gray scale images do not change
if a symmetric line profile is used instead of the mean observed
profile);
dark and light shades indicate less and more flux
compared to the mean profile, respectively.  In this representation,
in general, one can not safely discern an emission feature from an
absorption one moving through the line profile. If we interpret the
white feature in \ion{Ca}{2}~K as a chromospheric emission region,
then the corresponding dark feature simply results from the
normalization. On the other hand, if the dark feature is caused by an
absorption feature, then the remaining plot must be brighter simply to
fulfill the normalization condition.

Motivated by \citet{2005A&A...440.1105S} finding a potential
accretion spot in their Doppler images of MN~Lup, we inspected the
line profile variations shown in Fig.~\ref{fig:lines} for signatures
of a similar feature.  Such as spot should show up as 
emission, hence bright, feature in the chromospheric emission
profiles.  On the other hand, it would appear as \textit{dark} feature
in most photospheric lines: If the equivalent width of an absorption
line is approximately constant or decreases with temperature, then a
bright (i.e. hot) spot will cause a \textit{dip} (i.e. depression) in
the line profile, because its behavior is dominated by the
surrounding quasi-continuum \citep[e.g. Fig.~1
in][]{1983PASP...95..565V}.
The velocity where a rotationally modulated spot at latitude $\theta$
(defined to be $90^{\circ}$ at the pole and $0^{\circ}$ at the stellar
equator) and longitude $\phi_0$ ($\phi_0=0$ for spots that pass the
center of the stellar disk at phase $p=0$) is located, can be
calculated as
\begin{equation}  
v(p) = v\sin(i)\cos(\theta) \sin(2\pi(p-\phi_0))
\end{equation}
We did not find any combination of $\theta$ and
$\phi_0$ that fits slope and visibility of the observed bright feature.
As an illustration, the red sinusoidal line in
Fig.~\ref{fig:lines} traces approximately the bright feature in
\ion{Ca}{2}~K between phases 2.3 and 2.7 which could be due due to a
rotationally modulated hot spot fixed on the stellar surface and
rotating with it.  In this interpretation, the dark feature around the
same phases in \ion{Li}{1}, weaker in \ion{Ca}{1}, would represent the
photospheric signature of the tentative hot accretion spot.  However,
clearly the red sine does not trace the line profile variations
through \emph{all} observed phases: The \ion{Ca}{2}~K emission feature does
not continue before about phase 2.3, and, furthermore, it is not
visible two rotations earlier around phase 0.3.  The same observation
applies to the corresponding dark feature in the \ion{Li}{1} line.
Thus, the Ca K emission feature does not
seem to trace a hot spot on the stellar surface that survives the
observed stellar rotations.  We come to the same conclusion analyzing
the \ion{Ca}{2}~K emission feature passing the line profile center at
about phase 0.1. While the \ion{Ca}{1} line shows a fuzzy dark
structure at the same phases, this dark structure seems to passe the
line profile center at a later phase, not before phase 0.2.

We conclude that the chromospheric and photospheric line profiles in
our spectra do not show any clear signature of a hot, i.e. accretion,
spot rotating with the stellar surface.  On the other hand, our X-ray
analysis suggests ongoing weak accretion.  Unfortunately, it does not
yield any information on the location of this tentative accretion and
does not help in the above analysis.  If accretion is indeed present,
our optical data suggest that the corresponding photospheric and
chromospheric signature of the accretion funnel (a) is below our
current optical sensitivity, (b) smeared out on the surface or (c)
located so close to the visible pole that it does not lead to a
significant rotational modulation of the line profile. We cannot exclude any of
the above possibilities.  However, we consider option (c) rather
unlikely because our average line profiles do not show a pronounced
deformation near the line center which would be the spectroscopic
signature of this scenario.

\subsection{The H$\alpha$ line profile}
\label{sect:halpha}
\begin{figure}
\plotone{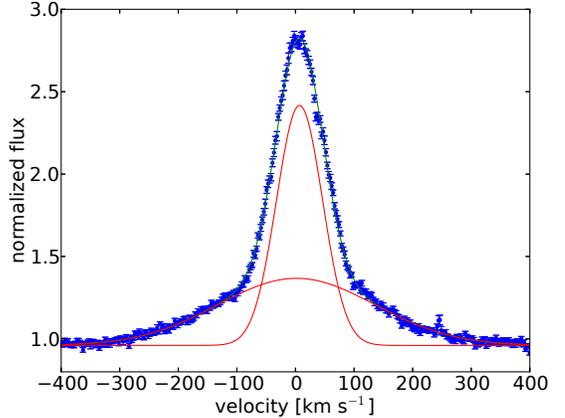}
\caption{Typical H$\alpha$ line profile with a fit of two Gaussian components. \label{fig:Ha}}
\end{figure}
H$\alpha$ and other Balmer lines can be used as an accretion tracer. Figure~\ref{fig:lc} shows the EW of the H$\alpha$ line together with the X-ray and UV light curves. The time resolution of the UVES data is too low to track the flares seen in the UV, but variability in the H$\alpha$ line exists on longer time scales. 

Figure~\ref{fig:Ha} shows a typical H$\alpha$ line profile. The line
can be well described by a narrow (FWHM $\approx 90$~km~s$^{-1}$) and
a wide (FWHM $\approx 300$~km~s$^{-1}$) Gaussian component. Both
components are almost centered: The narrow component shifts by less
then 1~km~s$^{-1}$ between the observations, the wide component by
25~km~s$^{-1}$. There are significant differences between the model at
+100 and \mbox{-100~km~s$^{-1}$}, indicating that the inner component
might differ from a Gaussian in the wings. \citet{2005A&A...440.1105S}
found a radial velocity of $4.4\pm2.0$~km~s$^{-1}$ and a peak-to-peak
variation of the line center of 9~km~s$^{-1}$. The peak-topeak variation is caused by the wide emission component. Active stars often show H$\alpha$ in emission with a narrow and symmetric profile \citep{2009A&A...504..461F}, while accreting CTTS typically show wide and asymmetric profiles with absorption and emission components both on the blue and on the red side of the line due to winds and accretion funnels, respectively.
Our observed H$\alpha$~EWs are similar to the value of 6.7~\AA{} found by \citet{1997A&AS..123..329K}.

\subsection{Forbidden optical emission lines}
We do not see any emission at the
position of the [\ion{O}{1}] 6364~\AA{} airglow line. A small peak,
which is however not significant above the noise, can be found at the
[\ion{O}{1}] 6300~\AA{} line. Other forbidden emission lines which are
commonly found in outflows from CTTS, e.g.\ [\ion{S}{2}] are not
visible in the spectrum of \citet{2005A&A...440.1105S} and this is
confirmed in our new data. The absence of any outflow tracers is 
consistent with the simple H$\alpha$ line profile.

\section{Discussion}
\label{sect:discussion}

\subsection{X-ray properties}
MN~Lup has been observed with \emph{ROSAT} in a survey and in three pointed observations with \emph{XMM-Newton}. 
MN~Lup is no longer embedded in a natal molecular cloud, so the interstellar absorption along the line-of-sight is low, but optical photometry requires reddening towards the stellar photosphere, which, assuming an interstellar $A_V/N_H$ ratio, predicts an absorbing column density about an order of magnitude larger than observed in our X-ray spectra.
We analyze both absorption scenarios in Sect.~\ref{sect:lowres} with the result that only the lower absorption column of $N_H = (2\pm1)\times10^{20}$~cm$^{-2}$ leads to physically consistent results.

Assuming a similar absorption, all measured X-ray luminosities from
\emph{ROSAT} and \emph{XMM-Newton} agree well. Like the H$\alpha$
emission, the X-ray luminosity does not show a long term trend for the last decade.
From the X-ray point of view MN~Lup is an active star. Its $L_X/L_{bol}$ ratio matches that of other saturated TTS or active MS stars.

In all spectra we see evidence of high temperatures, which can only be explained by coronal magnetic activity. Also, there is clear flaring in the lightcurves, both in 2010 and in 2011. In those flares, luminosity and temperature increase as expected for magnetic activity. This finding is supported by the IFIP effect in the elemental abundances, which is common in active stars. Alternatively, this pattern could be due to grain formation, where most of the Fe is bound in grains and remains in the disk, while more volatile gases like Ne remain in the gas phase and are accreted \citep{twhya}. However, no optically thick dust is observed around MN~Lup (see Sect.~\ref{sect:nodisk})

While the EPIC spectra are all compatible with stellar activity as expected from WTTS or young MS stars, the RGS data exhibit two phenomena, which point towards accretion: The low $f/i$ ratio in the \ion{O}{7} triplet and the soft excess. 

\subsection{Mass accretion}
\citet{2005A&A...440.1105S} argue that the Doppler image of MN~Lup can
best be explained, if a part of the photosphere is covered by the
accretion funnels. From their data they cannot determine the mass
accretion rate and they adopt $10^{-9} M_{\sun}$~yr$^{-1}$ as a fiducial value for their magnetic dipole
model, which would require a polar magnetic field of 15~kG on the
stellar surface. Smaller magnetic fields and accretion rates are possible down to $10^{-11} M_{\sun}$~yr$^{-1}$.

\citet{2003ApJ...592..282J} presented a correlation between accretion and the width of the H$\alpha$ line at 10\% of the maximum height. A similar formula was found by \citet{2003ApJ...582.1109W}. 
There is some discrepancy between the two groups about the borderline between accretion and purely chromospheric activity, as measured by the 10\% width of H$\alpha$. 
The first article sets this boundary at $200$~km~s$^{-1}$, the second at $270$~km~s$^{-1}$. In figure~\ref{fig:Ha} we measure exactly $270$~km~s$^{-1}$, which would correspond to accretion according to \citet{2003ApJ...592..282J}, while \citet{2003ApJ...582.1109W} still explain that as chromospheric activity.

\citet{2004A&A...424..603N} fit a relation between H$\alpha$ line
width at 10\% of the maximum and the accretion rate for a sample of
brown dwarfs accreting at very low rates. While the accretion process
could be similar, in the case of MN~Lup a much larger fraction of the
H$\alpha$ emission is due to chromospheric activity, as the X-ray
luminosity proves that it is an active star. Therefore, we consider
the value derived from this formula ($\dot M = 5\times 10^{-11}M_{\sun}$~yr$^{-1}$) as an upper limit. 

\citet{2011A&A...526A.104C} argue the H$\alpha$ line width at 10\% of
the maximum to be a good indicator of accretion, but a less reliable
way to calculate the accretion rate. Despite this, the upper limit
derived above compares well with the accretion rate of $2\times
10^{-11} M_{\sun}$~yr$^{-1}$ from our fit of X-ray accretion shock
models (Sect.~\ref{sect:shock}). This also rules out the scenario
where a significant part of the X-ray emission is buried as this
would lead to differences between the mass accretion rates measured
from H$\alpha$ and X-rays. We conclude that accretion with a very low rate takes place on MN~Lup.

One caveat is that the H$\alpha$ line in accreting stars is typically
asymmetric \citep{1998ApJ...492..743M,2003ApJ...592..266M} because the
emission partially originates in the accretion funnels, where we can
only observe the red-shifted accretion, since the funnels on the far
side of the star are blocked from view. Absorption of the H$\alpha$
emission from the accretion shock or in the chromosphere by a wind or
the accretion funnels also leads to asymmetric line profiles. However,
for very low accretion rates we expect only weak outflows and weak
accretion funnels, thus some geometric configurations could lead to a
symmetric H$\alpha$ line profile.

\subsection{Rotational modulation}
None of the X-ray or UV lightcurves in figure~\ref{fig:lc} shows any 
rotational modulation. Specifically, the X-ray lightcurve is flat in
2011 outside the large flare. On the other hand, in the Doppler images
of \citet{2005A&A...440.1105S} accretion spots can be seen at almost
any rotational phase with changes in the accretion shock area
observed and our analysis of the \ion{Ca}{2}~K, \ion{Li}{1}, and
\ion{Ca}{1} lines shows rotational modulation consistent with a feature at high stellar latitude.
While surveys of star forming regions rarely show a
correlation between rotationally modulated UV/optical and X-ray
lightcurves from CCD data \citep{2006ApJ...649..914S,2007A&A...468..379A}, \citet{2012ApJ...752..100A}
find rotational modulation in V4046~Sgr in the X-ray grating spectrum by adding
up the signal of those lines that are formed predominantly in the
accretion shock.
The absence of any significant X-ray modulation in the CCD spectra of MN~Lup is thus not sufficient to conclude that there is no accretion spot.

\subsection{Comparison to other TTS}
In many respects MN~Lup is similar to IM~Lup, a TTS star found in the same star forming region. IM~Lup is a transition case between CTTS and WTTS. Its inner disk is already evolved and, similar to MN~Lup, the corona shows hot components and an IFIP abundance pattern which point to magnetic activity. Just like MN~Lup, the X-ray luminosity of IM~Lup is close to the saturation limit \citep{2010A&A...519A..97G}. However, in IM~Lup the signal in the grating spectrum is too low to exclude either the high-density or the low-density limit of the $f/i$ ratio. IM~Lup was observed with \emph{Chandra}/HETG which offers a better wavelength resolution, but much lower effective area in the \ion{O}{7} triplet, so the density can only be observed in the \ion{Ne}{9} triplet. \ion{Ne}{9} is formed at a higher temperature, thus it has a stronger coronal contribution and much stronger shocks are required to cause a detectable signal.
On the other hand, IM~Lup has a more complex H$\alpha$ line with both emission and  absorption components. So, both stars seem to show accretion with a very low rate.

In comparison with the sample of CTTS of \citet{2011AN....332..448G},
which all have X-ray grating spectroscopy, MN~Lup shows similar
abundances, coronal temperatures and it has a comparable shock
speed. However, the density of the accreting plasma is at least an
order of magnitude lower than in TW~Hya, BP~Tau, V4046~Sgr and
MP~Mus. Surprisingly, due to the much larger area of the accretion
spot, the fitted mass accretion rate is similar. This could be partly
caused by a selection effect. TTS close to the end of their accretion
phase are the best to observe, since the absorption by the cloud is lower than in
younger objects. This could cause us to find predominately low, but
still detectable, accretion rates. It is possible, that the soft
emission in MN~Lup is a mixture of a denser accretion shock and a cool
corona. In this case the already low mass accretion rate would still
be overestimated by a factor of 2 (Sect~\ref{sect:gratings}).


\subsection{Accretion and the absence of a dust disk?}
\label{sect:nodisk}
Above, we argued that MN~Lup shows some signatures of weak, yet ongoing accretion at a rate of $2\times 10^{-11} M_{\sun}$~yr$^{-1}$. However, no IR excess was discovered in the \emph{Spitzer} c2d survey. This sets an upper limit on the fractional disk luminosity of $L_*/L_d < 9.1 \times 10^{-4}$ \citep{2010ApJ...724..835W}. 
Obviously, MN~Lup does not have a dusty disk, even the debris disk around $\beta$~Pic has a higher fractional luminosity.
How can these two observations be reconciled?

It is instructive to compare MN~Lup with \object{CVSO 224}, a M4 star
with a significantly lower X-ray luminosity \citep[$\log
L_X=29.3$,][]{2011AJ....141..127I}, but a few times larger mass
accretion rate  \citep[$7\times 10^{-11}
M_{\sun}$~yr$^{-1}$,][]{2008ApJ...689L.145E} estimated from a
H$\alpha$ line of similar width as in MN~Lup, but with a more complex
line profile. \citet{2008ApJ...689L.145E} detect a circumstellar disk
with an inner hole of 7~AU that contains small amounts of optically
thin dust.
In CVSO~224 the observed mass accretion rate is much lower than the
expected photo-evaporation rate given the X-ray and FUV
luminosity. Once photoevaporation becomes dominant, the inner disk
should clear on a very short time scale. The only way to explain mass
accretion in this context is a very massive outer disk that fuels a high accretion rate and due to planets or binary companions only a small fraction of this arrives on the star.
Based on this interpretation we postulate an as yet un-observed outer
disk for MN~Lup. However, the radius of the inner hole needs to be larger than in CVSO~224, since no excess beyond photospheric emission is seen at 24$\mu$m for MN~Lup. The c2d survey is most sensitive to micron-sized dust in the inner regions of an accretion disk; only mm data can reveal if a cool outer disk exists. If it does, then it could supply mass at a much higher rate than the observed stellar accretion rate. The inflow would be mostly stopped by photoevaporation as \citet{2012MNRAS.422.1880O} predict that the X-ray luminosity of MN~Lup can drive a photoevaporative wind with a mass loss rate of $5\times 10^{-9} M_{\sun}$~yr$^{-1}$ -- two orders of magnitude higher than the observed stellar mass accretion rate.
A second way to explain the mass for the accretion flow without an IR
excess are grains so large that they are invisible in the IR. Essentially, the inner disk would consist of sand and gas only. We can even speculate further that these grains may give rise to a secondary debris disk, when stellar activity, and thus the photoevaporation rate, decreases.

Alternatively, the observed phenomenae can be explained by MN~Lup's activity. The H$\alpha$ EW is well within the range of active stars \citep{1981ApJS...46..159W, 1990ApJS...74..891R}, but the wings of the line are unusually wide. Also, the X-ray luminosity and the IFIP pattern indicate an active star. On the other hand, the low-density limit for the \ion{O}{7} $f/i$-ratio is excluded on the 93\% confidence level with a lower limit of $3\times10^{10}$~cm$^{-3}$, which is above typical coronal densities in active stars \citep{2004ApJ...617..508T}.

Also, comparing the measured $N_H$ with the $A_V$ value from
\citet{2005A&A...440.1105S} we find a low $N_H/A_V$ ratio compared with the ISM.
This could be caused by a lower gas-to-dust ratio, though there is no reason to expect that material around MN Lup should be dust-enhanced. On the contrary, the stellar radiation should evaporate the dust. 
One explanation could be that the reddening is very local. This would fit the suggestions by \citet{2005A&A...440.1105S} that part of the reddening is due to the accretion columns. Conceivably, different X-ray emission components could be effected differently: While the corona is directly observable, the accretion shock might be hidden behind the accretion stream and in X-rays we would measure an average of the different $N_H$ values. However, the signal-to-noise ratio of the X-ray spectrum is insufficient to fit multiple absorption components.
On the other hand, the $N_H/A_V$ ratio is not only influenced by the gas-to-dust ratio, but also by the dust composition. Grain coagulation can increase the reddening per hydrogen atom and thus cause lower $N_H/A_V$ ratios. While these changes are most obvious around the silicate feature in the IR, they pertain to optical wavelengths as well \citep{2011A&A...532A..43O}.
Varying $N_H/A_V$ ratios are observed for different molecular clouds \citep[see e.g. the discussion in ][]{2012AJ....144..101G}, so deviations from the interstellar ratio are certainly consistent with the presence of circumstellar material.


\section{Summary and conclusion}
\label{sect:summary}
We observed MN~Lup with \emph{XMM-Newton} and \emph{VLT}/UVES with
some of the data taken simultaneously. There are three short ($\approx
1$~ks) flares seen in the UV and the X-rays as well as
longer periods of enhanced X-ray activity in 2010 and towards the end
of the 2011 observations with a harder spectral signature. 
The short UV and X-ray bursts have all characteristics of solar
``white-light'' flares.

The X-ray spectrum is fitted with three components, the
softest of which is best described by an accretion shock model. The abundances
follow an IFIP pattern and the ratio $L_X/L_{bol}= -2.7$
indicates that MN~Lup is an X-ray active star. 
From the \ion{O}{7} triplet in the grating spectrum we derive a lower
limit on the density of $n_e > 3\times 10^{10}$~cm$^{-3}$, which is
higher than normally seen in active stars on the 93\% confidence level. This could be the signature
of an accretion shock with a low mass accretion rate around
$2\times10^{-11}$~M$_\sun$~yr$^{-1}$. The H$\alpha$ line consists of a
narrow and a wide component. While the narrow line is due to
chromospheric activity, the wide component again is unusual for active
stars on the MS and hints at accretion activity. Analysis of
rotationally modulated line profiles in \ion{Ca}{2}~K, \ion{Li}{1},
and \ion{Ca}{1} indicates one or more features at high stellar latitude. However, no dust disk that could fuel the accretion was found
around MN~Lup in a \emph{Spitzer} survey.
In the H$\alpha$ line we see a prominence in absorption for the first
time in any TTS.

MN~Lup is a unique case in the evolution from CTTS to WTTS. While some non-accreting
WTTS still retain detectable disks, MN~Lup is the only known case of
an accreting star without an IR excess. 
We propose that MN~Lup has lost its optically thick
dust only recently. Since photoevaporation clears the inner disk in
short time, the continued accretion requires a mass reservoir. No
IR excess is detected, so this mass must either be located at large disk radii ($>10$~AU)
or grain growth has formed an inner disk of particles with sizes of a
few mm or above, that do not radiate in the IR. In either case, the
disk of MN~Lup would be unlike any other known accreting TTS and might
well present the very last stage where accretion can be
seen. Its radial structure (either a large inner hole or very evolved
dust) would be consistent with the recent formation of planetesimals
or planets.

It is a lucky coincidence that we could study MN~Lup in this
detail. Without the archival Doppler imaging, no X-ray observation
long enough to measure the $f/i$ ratio would have been performed and
thus it is unlikely that more objects with these characteristics can
be found. The next step to test our ideas about the disk of MN~Lup is
to search for emission from cool dust with mm observations.

\acknowledgements 
Based on observations obtained with XMM-Newton, an ESA science mission with instruments and contributions directly funded by ESA Member States and NASA and the ESO VLT. HMG was supported by the National Aeronautics and Space Administration under Grant No. NNX11AD12G issued through the Astrophysics Data Analysis Program.

{\it Facilities:} \facility{XMM} \facility{VLT:Kueyen}


%
\bibliographystyle{../AAStex/astronat/apj/apj}
\bibliography{../articles}


\end{document}